\pdfoutput=1
\documentclass{raa_rb}
\usepackage{graphicx,times}
\usepackage{natbib}
\usepackage{amssymb,amsmath}
\usepackage{subcaption}
\bibpunct{(}{)}{;}{a}{}{,}
\usepackage{hyperref}

\begin{document}

\newcommand{\BI}{\textit{Breakthrough Initiatives}}
\newcommand{\BLI}{\textit{Breakthrough Listen Initiative}}
\newcommand{\BL}{\textit{Breakthrough Listen}}
\newcommand{\SKA}{\textit{Square Kilometre Array}}
\newcommand{\MK}{MeerKAT}
\newcommand{\gmrt}{uGMRT}
\newcommand{\VLA}{\textit{Expanded Very Large Array}}
\newcommand{\Parkes}{\textit{Parkes Observatory}}
\newcommand{\MWA}{\textit{Murchison Widefield Array}}
\newcommand{\LOFAR}{\textit{LOw Frequency ARray}}
\newcommand{\ATA}{\textit{Allen Telescope Array}}
\newcommand{\APF}{\textit{Automated Planet Finder}}
\newcommand{\subtot}[1]{{\textcolor{Cerulean}{#1}}}
\newcommand{\subsubtot}[1]{\textit{\textcolor{OliveGreen}{#1}}}
\newcommand{\tot}[1]{{\textcolor{NavyBlue}{#1}}}
\newcommand{\tseti}{{\itshape turbo}SETI}
\newcommand{\UCB}{Department of Astronomy,  University of California Berkeley, Berkeley CA 94720}
\newcommand{\SSL}{Space Sciences Laboratory, University of California, Berkeley, Berkeley CA 94720}
\newcommand{\UCBEECS}{Department of Electrical Engineering and Computer Sciences,  University of California Berkeley, Berkeley CA 94720}
\newcommand{\SWIN}{Centre for Astrophysics \& Supercomputing, Swinburne University of Technology, Hawthorn, VIC 3122, Australia}
\newcommand{\GBT}{Green Bank Observatory,  West Virginia, 24944, USA}
\newcommand{\OXF}{Astronomy Department, University of Oxford, Keble Rd, Oxford, OX13RH, United Kingdom}
\newcommand{\NIJ}{Department of Astrophysics/IMAPP,Radboud University, Nijmegen, Netherlands}
\newcommand{\ATNF}{Australia Telescope National Facility, CSIRO, PO Box 76, Epping, NSW 1710, Australia}
\newcommand{\HOU}{Hellenic Open University, School of Science \& Technology, Parodos Aristotelous, Perivola Patron, Greece}
\newcommand{\USQ}{University of Southern Queensland, Toowoomba, QLD 4350, Australia}
\newcommand{\PENN}{Department of Astronomy and Astrophysics, Pennsylvania State University, University Park PA 16802}
\newcommand{\BTI}{Breakthrough Initiatives, NASA Research Park, Moffett Field CA 94043 USA}
\newcommand{\SETI}{SETI Institute, Mountain View, California}
\newcommand{\KZA}{University of Malta, Institute of Space Sciences and Astronomy}
\newcommand{\ICRAR}{International Centre for Radio Astronomy Research, Curtin University, Australia}
\newcommand{\NAOC}{CAS Key Laboratory of FAST, NAOC, Chinese Academy of Sciences (CAS) Beijing,100101, China}
\newcommand{\CASS}{University of Chinese Academy of Sciences, Beijing 100049, People’s Republic of China}
\newcommand{\XAO}{Xinjiang Astronomical Observatory, CAS, 150, Science 1-Street, Urumqi, Xinjiang 830011, China}
\newcommand{\BNU}{Department of Astronomy, Beijing Normal University, Beijing 100875, China}
\newcommand{\SDU}{Institute of Astronomical Science, Dezhou University, Dezhou 253023, China}

\title{Opportunities to Search for Extra-Terrestrial Intelligence with the
   Five-hundred-meter Aperture Spherical radio Telescope}

 \volnopage{ {\bf 20XX} Vol.\ {\bf X} No. {\bf XX}, 000--000}
   \setcounter{page}{1}

   \author{Di Li\inst{1,2},
   Vishal Gajjar\inst{3},
   Pei Wang\inst{1},
   Andrew Siemion\inst{3,4,5,6},
   Zhisong Zhang\inst{1,2},
   Haiyan Zhang\inst{1,2},
   Youling Yue\inst{1,2},
   Yan Zhu\inst{1,2},
   Chengjin Jin\inst{1,2},
   Shiyu Li\inst{2},
   %FAST team\inst{1,2},
   %Breakthrough Listen team\inst{3}
   Sabrina Berger\inst{3},
   Bryan Brzycki\inst{3},
   Jeff Cobb\inst{7},
   Steve Croft\inst{3},
   Daniel Czech\inst{3},
   David DeBoer\inst{3},
   Julia DeMarines\inst{3},
   Jamie Drew\inst{8},
   J. Emilio Enriquez\inst{3,5},
   Nectaria Gizani\inst{3,9},
   Eric J. Korpela\inst{7},
   Howard Isaacson\inst{3,10},
   Matthew Lebofsky\inst{3},
   Brian Lacki\inst{3},
   David H.\ E.\ MacMahon\inst{3},
   Morgan Nanez\inst{3},
   Chenhui Niu\inst{1,2},
   Xin Pei\inst{11,2},
   Danny C.\ Price\inst{3,12},
   Dan Werthimer\inst{3},
   Pete Worden\inst{8},
   Yunfan Gerry Zhang\inst{3},
   Tong-Jie Zhang\inst{13,14}, 
   and FAST Collaboration
   }
 \institute{{\NAOC}; {\it dili@nao.cas.cn} %1
   \and
   {\CASS}%2
   \and
  {\UCB}; {\it vishalg@berkeley.edu}%3
	\and
  {\SETI}%4	
    \and
  {\NIJ}%5
    \and
  {\KZA}%6
    \and
  {\SSL}%7
  \and
  {\BTI}%8
  \and
  {\HOU}%9
  \and
  {\USQ}%10
    \and
  {\XAO}%11
    \and
  {\SWIN}%12
    \and
  {\BNU}%13
    \and
  {\SDU}%14 
%	  Center for Astrophysics, University of Science and Technology of China, Hefei 230026, China\\
}

\abstract{
The discovery of ubiquitous habitable extrasolar planets, combined with revolutionary advances in instrumentation and observational capabilities, has ushered in a renaissance in the search for extra-terrestrial intelligence (SETI).  Large scale SETI activities are now underway at numerous international facilities. The Five-hundred-meter Aperture Spherical radio Telescope (FAST) is the largest single-aperture radio telescope in the world, well positioned to conduct sensitive searches for radio emission indicative of exo-intelligence. SETI is one of the five key science goals specified in the original FAST project plan.  A collaboration with the \BL\ Initiative has been initiated in 2016 with a joint statement signed both by Dr. Jun Yan, the then director of the National Astronomical Observatories, Chinese Academy of Sciences (NAOC), and Dr. Peter Worden, the Chairman of the Breakthrough Prize Foundation. In this paper, we highlight some of the unique features of FAST that will allow for novel SETI observations.  We identify and describe three different signal types indicative of a technological source, namely, narrow-band, wide-band artificially dispersed, and modulated signals. We here propose observations with FAST to achieve sensitivities never before explored.
}

   \authorrunning{Li, D. et al. }            %author_head in even pages
   \titlerunning{Opportunities for SETI with FAST}  % title_head in odd pages
   \maketitle

%________________________________________________ sections below
%
\section{Introduction}           %% first-level sections will be auto-capitalized
\label{sect:intro}
%Motivation for SETI
%JD: COMMENT: "...to be capable of communicating [emitting?] using electromagnetic..."
The search for life beyond Earth seeks to answer one of the most profound questions regarding human being's place in the universe $-$ Are we alone?  Recent discoveries of thousands of exoplanets, including many Earth-like planets \citep{2014PASP..126..398H,2015ApJ...807...45D}, generate abundant targets of  interest. It is possible that some fraction of these planets host life sufficiently advanced to be capable of communicating using electromagnetic waves. Coherent radio emission is commonly produced by our technology for various applications. Moreover, radio waves are also energetically cheap to produce and can convey information at maximum speed across large interstellar distances. \cite{1959Natur.184..844C} speculated that frequencies near 1420 MHz (the Hydrogen line) are particularly well suited for interstellar communication.  Later it was suggested that frequencies between the hyperfine hydrogen transition and the $\Lambda$-doubling OH lines (1400 MHz $\sim$ 1700 MHz), could be considered a ``cosmic water-hole" \citep{NASA:2003p185}, where intelligent species might transmit a deliberate beacon to other technologically advanced species\footnote{We note that our implicit definition of advanced species only considers civilizations who have developed radio communication capabilities.}.

%\section{Past and on-going SETI}
%\label{sect:surveys}
%1. Some of the famous SETI experiments and their outcomes (may be a Table of summary).
Radio astronomy has long played a prominent role in SETI. Large single dish radio telescopes, with their enormous collecting area, flexibility of operation, and large beams relative to interferometers, are ideal for both targeted and large-area SETI surveys. \cite{Drake:1961bv} conducted some of the earliest SETI experiments near 1420 MHz towards two stars using the National Radio Astronomy Observatory's (NRAO) 26-meter radio antenna. Other radio telescopes such as the Arecibo radio telescope (Arecibo), Parkes radio telescope (Parkes), NRAO's 91-meter, 300-feet, and 140-feet dishes have also been used for SETI experiments \citep{Horowitz:1993p1523,1986Icar...67..525H,Tarter:1980p1516,Verschuur1973}.
Most of these early studies were limited to only a few stars. One of the largest SETI experiments of the 20th century, {\itshape Project Pheonix}, was conducted from the Arecibo, Parkes, and NRAO's 43-m telescopes, and surveyed around 1,000 nearby stars \citep{Tarter:1996jf,1998AcAau..42..635D,BACKUS1998,  2000ASPC..213..451C,2002ASPC..278..525B}. Later studies by \cite{2013ApJ...767...94S} used the 100-meter Robert C. Byrd Green Bank Radio Telescope ({\itshape hearafter} GBT) for a targeted search towards 86 stars in the Kepler field. Most recently, \cite{2017ApJ...849..104E} have published the most comprehensive targeted radio SETI survey of 692 nearby stars, also using the GBT. It should be noted that, in congregate, all the several dozen significant radio SETI spanning over the last six decades, have explored only a fraction of the multidimensional parameter space of potential signals \citep{Tarter:2003p266,2018AJ....156..260W}.

The \BLI\ (BLI) is a US \$100M 10$-$year effort to conduct the most sensitive, comprehensive, and intensive search for advanced life on other worlds \citep{2017AcAau.139...98W,Isaacson:2017ib}.  BLI is currently utilizing dedicated time on three telescopes, including the GBT \citep{MacMahon:2017we} and  Parkes \citep{Price:2018bv} operating at radio wavelengths and the optical \APF\ \citep{Lipman:2018wv}. The BLI team has leveraged both standard and bespoke tools to construct a flexible software stack to search data from these and other facilities for signals of interest.
%\footnote{\url{https://breakthroughinitiatives.org/news/6}}.

FAST is the largest single aperture radio telescope in the world \citep{rendong:2006,rendong:2011,li:2016}. With the newly cryogenically-cooled FAST L-band  Array of 19-beams (FLAN: Li et al.\ 2018), FAST is poised to become one of the most sensitive and efficient instruments  for radio SETI experiments. Since the early days of its conception, \cite{rendong:2000} and \cite{bopeng:2000} have indicated that FAST will be a leading facility to search for signals of extra-terrestrial intelligence (ETI). \cite{rendong:2000} also highlighted that a dedicated SETI survey with FAST will be 2.5 times more sensitive and will be able to cover five times more stars than the aforementioned {\itshape Project Pheonix}. FAST surveys will be complimentary to sensitive SETI experiments being carried out by BLI. In 2016, BLI signed a Memorandum of Understanding with the National Astronomical Observatories, Chinese Academy of Sciences (NAOC) for future collaboration with FAST.

We outline here novel SETI experiments possible with FAST and quantify their expected outcomes based on test observations. Section \ref{sect:setiWfast} highlights two targeted SETI experiments, namely, a survey of nearby galaxies and nearby stars with newly-discovered exoplanets. In order to conduct these surveys, a dedicated instrument is required which can capture baseband raw voltages to acquire the data products for various signal searches.  The \BL\ team has developed tools to capture baseband voltages into the Green Bank Ultimate Pulsar Processing Instrument (GUPPI \footnote{The GUPPI support guide can be found @ \url{https://safe.nrao.edu/wiki/bin/view/CICADA/GUPPISupportGuide}} ) raw format and then convert them to the various spectral and temporal resolutions during the offline processing \citep{mpl+18,lcs+19}. More details about this backend are also discussed in Section \ref{sect:system}.
%\del{We are in a new era, when a broad range of signal types can be investigated.} 
In Section \ref{sect:signalsearches}, we discuss three potential signal types, which are likely to provide a new window for radio SETI experiments. In Section \ref{sect:limits}, we estimate the sensitivities of these FAST SETI surveys. We also describe a FAST experiment with just a few hours to potentially place the most stringent limits on the presence of artificial transmitters within its operable frequency range. 

\section{FAST SETI System}
\label{sect:system}

FAST has an active, segmented primary surface of 500 meters in diameter, with a maximum effective aperture of 300 meters. The receiver cabin is driven by 6 cables connecting to mechanic drives through 6 towers. Pointing and tracking can be accomplished by reforming the primary and drive the focal cabin on the curved aperture plane. Such a concept provides access to much larger region of the sky ($-14^\circ$ to $+66^\circ$ in declination) compared to Arecibo \citep{Li:2018}. The FAST L-band  Array of 19-beams is the largest of its kind, comparing to the 7 beams of Arecibo and 13 beams of Parkes, which provides both unprecedented survey speed and efficiency in discriminating radio frequency interference (RFI) for SETI.
FLAN has a T$_{sys}$ between 18-24 K, which, combined with the effective aperture, amounts to a gain of 16 K/Jy, which will enable 2.5 times more sensitive SETI surveys than Arecibo (10 K/Jy gain and 30 K T$_{sys}$)\footnote{\url{ http://www.naic.edu/\%7Eastro/RXstatus/rcvrtabz.shtml}} at similar frequencies.
\begin{figure}[h]
    \centering
    \includegraphics[scale=0.5]{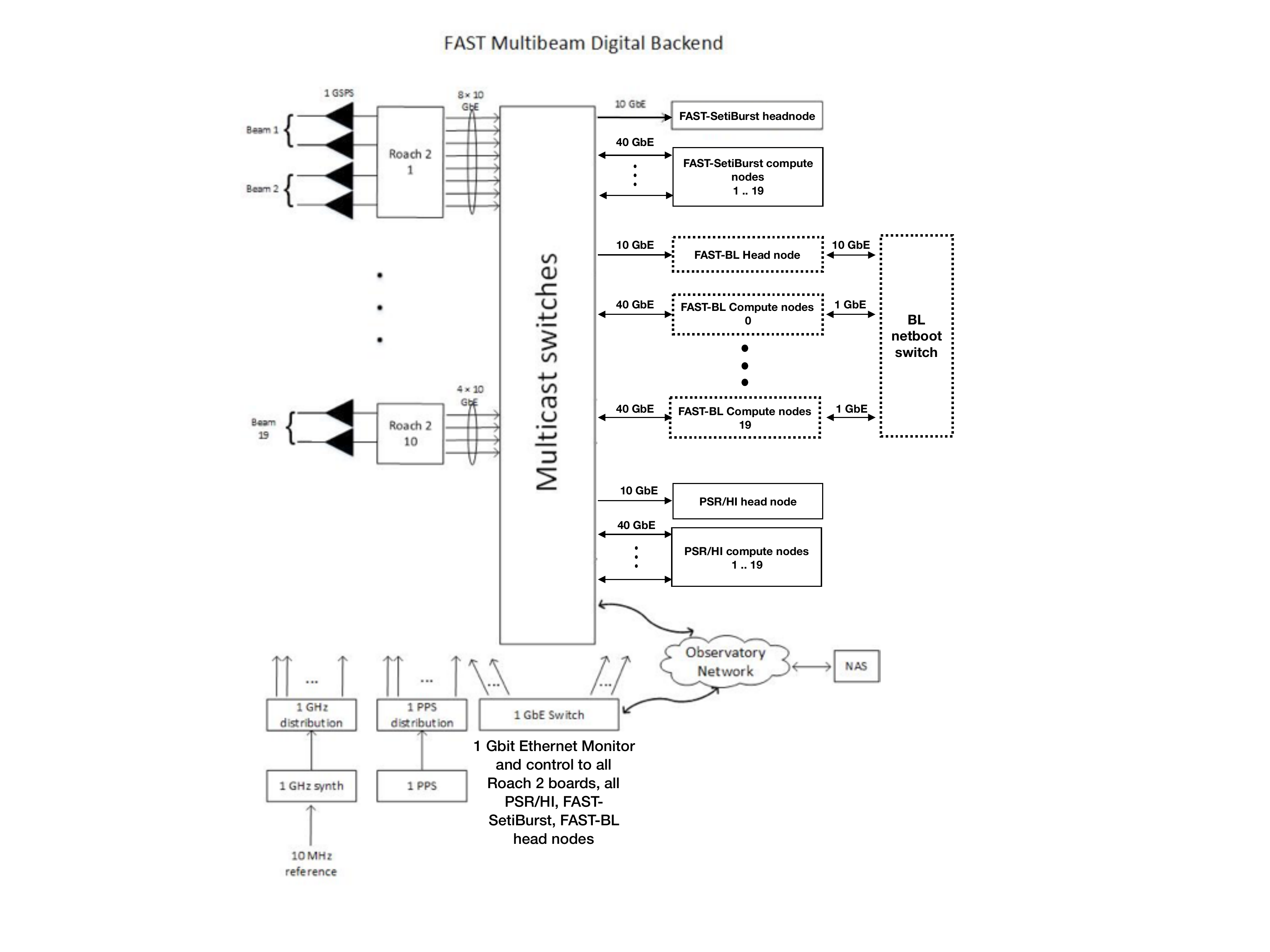}
    \caption{The schematic diagram of the FAST multibeam digital backend for the FAST L-band Array of 19-beams (FLAN). Note that the RF signals transmitted over fiber from the dome are digitized simultaneously, which allows for a potential correlation between beams. There are other multiple observatory based backends already installed with the multicast switches, shown with solid lines. One possible configuration for connecting a dedicated compute cluster in collaboration with the \BL\ team is shown with dotted boxes. This computing cluster could be consist of one headnode used for system monitor and metadata collection, 19-compute nodes with GPUs for recording and offline processing, and a dedicated switch for interconnectivity between these computing nodes.}
    \label{fig:BL_backend_nodes}
\end{figure}{}

As one of the major FAST surveys, the Commensal Radio Astronomy FAST Survey (CRAFTS  \citet{Li:2018}) will utilize FAST's 19-beam receiver to conduct searches for pulsars and HI galaxies, HI imaging, and fast radio bursts (FRBs). An additional backend, namely named {\itshape FAST-SetiBurst} (Figure \ref{fig:BL_backend_nodes}), has been installed for a commensal SETI -- which can record thresholded spectra -- focusing on narrow-band signals. This instrument is similar in design and operation to the seven beam backend at the Arecibo \citep{2017ApJS..228...21C}.  The SETI section of the {\itshape FAST-SetiBurst} instrument consists of 38 ultra high resolution FPGA/GPU based spectrometers for each of the 19 beams of the multibeam receiver (for both polariazations). Each 250 million channel spectrometer has 4 Hz spectral resolution and covers a 450 MHz band, from 1025 to 1475 MHz. The spectrometers search in real-time for narrow band signals and output files that contain a list of narrow band signals with their frequency, Julian time, power, position in the sky as well as other meta information. The {\itshape FAST-SetiBurst} instrument also continuously records raw voltage streams from all 19 beams and both polarizations (38 signals at 500 MHz bandwidth each, totaling 38 billion samples per second).  This instrument and associated software are open source and available for use by all FAST observers for SETI and FRB science related projects. A more detailed description of this FAST specific backend will be presented elsewhere in future publications (Zhang et al. 2020, accepted ApJ). 

It should be noted that FAST-SetiBurst can not do coherent search for drifting narrow-band signals as it only stores thresholded spectra. Moreover, for more complicated classes of signal searches, baseband raw-voltages are required to be utilized and coherent searches need to be performed. Section \ref{sect:signalsearches} outlines some of these searches that are possible to carry out if raw-voltages can be acquired and processed. Figure \ref{fig:BL_backend_nodes} also shows prospective GPU equipped compute nodes that can be connected to the existing multicast switch network to capture coarsely channelized baseband raw-voltage spectra from the ROACH-II\footnote{\url{https://casper.ssl.berkeley.edu/wiki/ROACH-2_Revision_2}} boards and convert them to GUPPI formatted raw-voltages for offline processing. The \BL\ team has deployed such computing clusters at the Green Bank Telescope \citep{mpl+18} and Parkes Telescope \citep{pml+18}. We refer readers to these references for further details on the architecture of such a computing cluster.

%\begin{figure}[h]
% \centering
%    \includegraphics[scale=0.5]{fast-backend.pdf}
%    \caption{{\bf The schematic diagram of the SETI backend for the  FAST L-band % Array of 19-beams (FLAN). Note that the RF signals transmitted over fiber from %the dome are digitized simultaneously, which allow for potential correlation %between beams. The SETI backend utilizes the same hardware as the FRB backend. %}}.
%    \label{fig:backend}
%\end{figure}

\begin{figure}[h]
 \centering
    \includegraphics[scale=0.5]{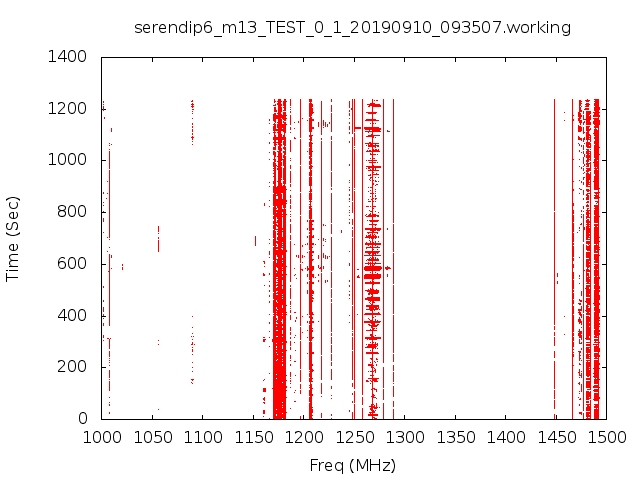}
    \caption{The "waterfall-plot" is the preliminary result using the {\itshape FAST-SetiBurst} backend. The red points in the figure represent signals with S/N greater than 30 that have been extracted by the pipeline, containing mostly narrow-band and broad-band RFI. We identify a relatively clean window between 1300-1450 MHz}.
    \label{fig:waterfall}
\end{figure}

To demonstrate capabilities of the FAST telescope for SETI, we conducted preliminary observations with the {\itshape FAST-SetiBurst} with five minutes tracking observations toward GJ273b on September 10th, 2019. GJ273b is one of the closest Earth-size planet inside the habitable zone of an M-dwarf star \citep{tjb19}, making it one of the most interesting target for deep dedicated SETI experiments. The preliminary result from the FAST-SetiBurst  backend is shown in  Figure~\ref{fig:waterfall}.  The data were taken with 4 Hz channel width, 0.25 s integration time, and processed with a 30 S/N threshold cutoff. The noise level is consistent with the expectation from the radiometer equation for a system of 20 K and effective gain of 16 K/Jy. As these are thresholded spectra, they can not be added for the entire length observing duration like in case of \cite{2017ApJ...849..104E}. Thus, with $t_{obs}$ of 0.25 seconds, for GJ273b located at 12 light-years, estimated EIRP limit see is around 7$\times$10$^{10}$ W.

\section{SETI Surveys with FAST}
\label{sect:setiWfast}
% Importance of FAST (A table of SETI related parameters)
\cite{Kardashev:1964} proposed a classification scheme for technological civilizations based on their energy utilization capabilities. A Kardashev Type I civilization is defined as one that can harness all the stellar energy falling on their planet (around 10$^{17}$ W for an Earth-like planet around a Sun-like star) and a Type II civilization  as one that can harness the entirety of the energy produced by their star (around 10$^{26}$ W for the Sun). The Kardashev Type III would be capable of harnessing all the energy produced by all the stars in a galaxy, around 10$^{36}$ W for a Milky Way-like galaxy. The likelihood of the existence of such  civilizations among a given number of stars might be sparse, thus, a comprehensive search for Type II and Type III civilizations should be conducted towards a large number of stars. 
We thus quantify the expected performance of two  FAST SETI surveys in terms of Kardashev types, namely a deep blind search toward the Andromeda galaxy and a  targeted search toward TESS stars with exoplanets.

\subsection{Andromeda system (M31) with FLAN}

\begin{figure}[h]
    \centering
    \includegraphics[scale=0.3]{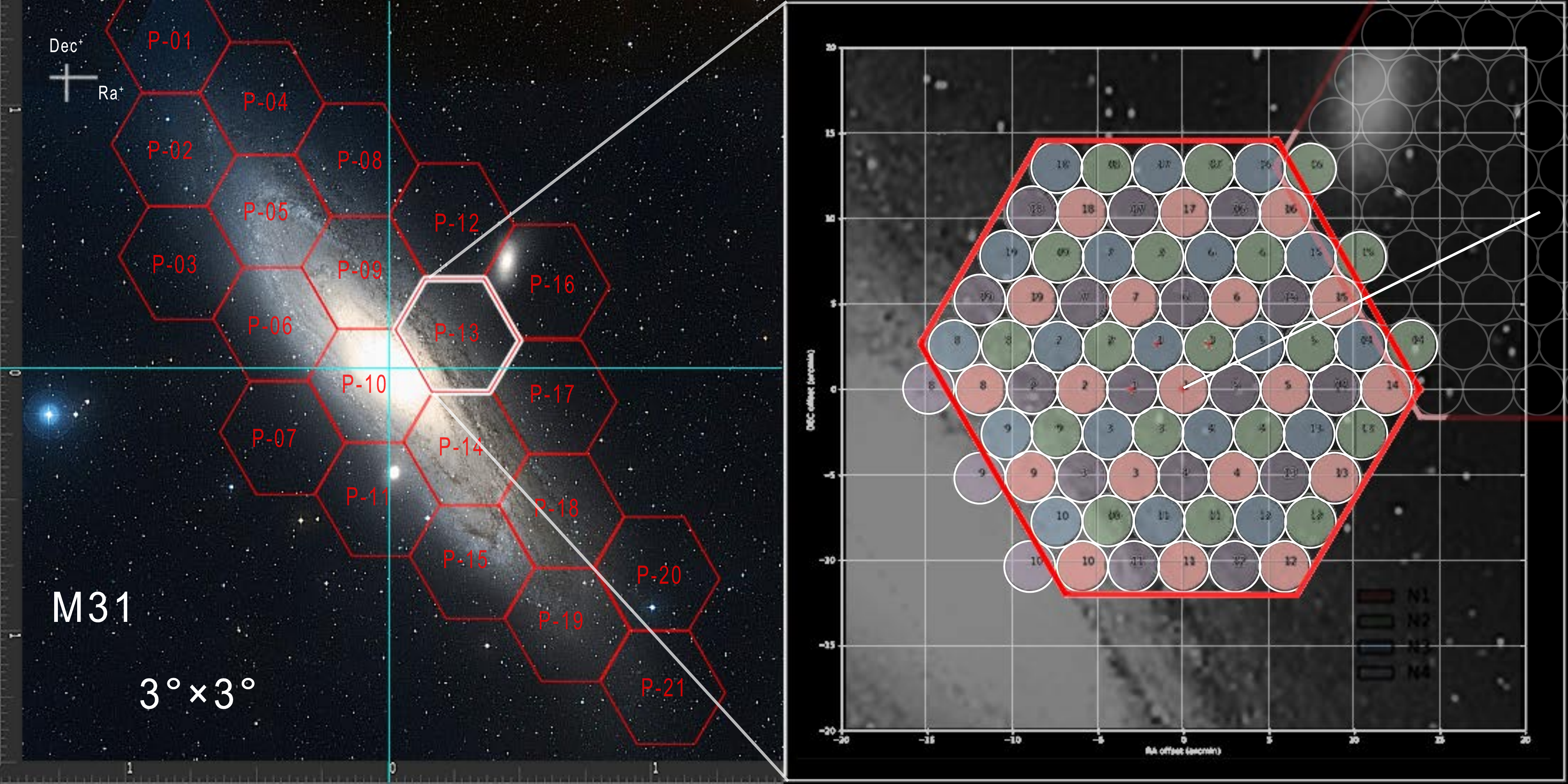}
    \caption{Proposed RFI-resistant SETI towards the Andromeda Galaxy using the FAST 19-beam receiver, surveying one trillion stars. The inset on the bottom-left corner shows four pointings filling a hexagon, with each pointing shown in a different color. A total of 21 such hexagon tiles are required to cover the Andromeda galaxy. The image file was taken from {\itshape Digital sky survey} archival data (www.archive.eso.org/dss/dss)}.
    \label{fig:FAST19beam}
\end{figure}

Nearby galaxies such as Andromeda (M31) and M33 are ideal targets for SETI surveys aiming for very advanced civilizations. The radio interferometers, such as the Very Large Arrray (VLA) and MeerKAT, provide decent sensitivity and spatial dynamic range for wide-field SETI towards such nearby galaxies. The data rates of such interferometric surveys, however, will remain extremely challenging for the near future\footnote{The \BLI\ is in the process of deploying a state-of-the-art 128-node computing cluster at MeerKAT, which will be one of the largest backends ever deployed for radio astronomy, to mitigate these challenges.}. For example, the recent VLA SETI experiments by \cite{2017AJ....153..110G} towards M31 and M33 were limited to a very small spectral window ($\sim$1 MHz) around 1420 MHz.
%Other larger radio telescopes such the GBT can provide necessary sensitivities, however, due to the existing single beam at L-band (8\' beam), it will require relatively larger number of pointings ($>$100) and significantly larger number of observing hours.
Centered around 1250 MHz, the FLAN provides unprecedented sensitivity over 400 MHz bandwidth . Figure \ref{fig:FAST19beam} highlights the FLAN's tiling scheme, which forms a hexagon of about 26' across with four pointings and  covers the entire Andromeda galaxy with 21 such hexagons. A dedicated FAST SETI survey of M31with 10 minutes  per pointing (see table 1 and Section \ref{sect:limits} for sensitivity discussion) will be able to identify any continuous and isotropic transmitters from Type II and Type III civilizations among one trillion stars in the Andromeda galaxy.

\subsection{Survey TESS stars across 70 MHz to 3 GHz}
\label{sect:TESS}
Over the last two decades, more than 3,000 exoplanets have been
discovered,  among which several dozen have been confirmed to be inside
%\footnote{\url{www.exoplanetarchive.ipac.caltech.edu/docs/counts_detail.html} }.
the putative habitable zone\footnote{\url{www.phl.upr.edu/projects/habitable-exoplanets-catalog}}. According to recent estimates, the average number of planets per star is greater than one \citep{Zink:2019eb}.
The Transiting Exoplanet Survey Satellite (TESS) was launched in April of 2018 first surveying the southern sky before turning to the northern sky in the summer of 2019. TESS fully covers its 24$^{\circ}\times$96$^{\circ}$ field-of-view every 30 minutes, while measuring 200,000 bright stars on a two minute cadence in search of Earth-sized planets. Such a large field of view allows for 80\% coverage of the entire sky in its first two years \citep{2018AJ....156..102S}.
%Each pointing lasts 27 days, and overlapping pointings allow for the monitoring of the southern and northern ecliptic poles for 300 days. Compared to the Kepler spacecraft, TESS will survey a sky area 400 times larger, and the typical detected exoplanets  will be 10 times brighter than those detected by Kepler.
%The primary science goal of TESS is to identify candidates for exoplanet atmospheric study. TESS is
%focused on bright stars, many nearby that are conducive to in transit atmospheric study by JWST.
Pre-flight estimates of the planet yield suggest that TESS will find 1250 planets, 250 of which are at most twice the size of Earth around likely bright dwarf stars. As many as 10,000 planets could be found in the full-frame images around fainter stars
\citep{barclay:2018}. These stars are less likely to have been studied with deep observations in earlier SETI experiments. Thus, many of them will be ideal targets for a dedicated SETI project with FAST.

\begin{table}[h]
    \centering
    \begin{tabular}{cccccc}
     \hline
       Receiver Name  & RF Band (MHz) & Number of beams & Polarization & T$_{sys}$ K & EIRP limits (W) \\
       \hline
       \hline
            A1        & 70-140   &     1           &  Circular   &  1000  & 9.5$\times$10$^{12}$ \\
            A2        & 140-280  &     1           &  Circular   &  400   & 3.8$\times$10$^{12}$ \\
            A3        & 560-1120 &     1           &  Circular   &  150   & 1.4$\times$10$^{12}$ \\
            A4        & 1100-1900 &     1           &  Circular  &  60    & 5.7$\times$10$^{11}$ \\
            {\bf FLAN}    & 1050-1450 &     19           &  Linear   &  20    & 1.9$\times$10$^{11}$ \\
            A5        & 2000-3000 &     1            &  Circular &  25    & 2.4$\times$10$^{11}$ \\
        \hline
    \end{tabular}
    \caption{A collection of receivers expected to be available with FAST. EIRP limits are calculated for putative 1 Hz bandwidth signals at a distance of 200 light years observed with 5 minute integration (see Section \ref{sect:limits}.) The T$_{sys}$ values are taken from \cite{rendong:2011}.
    %{\bf except for FLAN, which has been updated with real measurements}.
    }
    \label{tab:all_receivers}
\end{table}

We expect to use all  receivers listed in Table \ref{tab:all_receivers} (A1, A2, A3, A4, and A5) from 70 MHz to 3 GHz to conduct a deep and comprehensive search towards a subset of stars with exoplanet systems discovered by TESS. As indicated in Table \ref{tab:all_receivers}, such a survey would provide some of the most constraining limits (see Section \ref{sect:limits}) on possible narrow-band transmission yet achieved in radio SETI. It should be noted that although these limits are calculated for narrow-band signals, the proposed observations would provide correspondingly robust limits for other signal types.

\section{Signal searches}
\label{sect:signalsearches}
\subsection{Narrow-band signals}
Narrow-band ($\sim$ Hz) radio signals are one of the most common signal types aimed for by radio SETI.  Ubiquitous in early terrestrial communication systems, such signals can be produced with relatively low energy and transverse the interstellar medium easily. They can be readily distinguished from natural astrophysical sources. These signals could either be transmitted intentionally or arise as leakage from extrasolar technologies. The apparent frequency of a distant narrow-band transmitter is expected to exhibit Doppler drift due to the relative motion between the transmitter and receiver. For a radio signal transmitted at rest relative to the Earth's barycenter at frequency, $\nu_0$, can be expressed as
\begin{equation}
    \dot{\nu}_{Doppler} = \frac{\omega^2{R}{\nu_0}}{c}\>\>,
\end{equation}
where $\omega$ is the angular velocity, $R$ is the radius of Earth, and
$c$ is the speed of light.  For a transmitter operating at 1400 MHz,
the frequency drift rate is $\sim$0.14 Hz/sec.  While the motion of the Earth is well known and can be exactly removed, the intrinsic or rotational/orbital drift of an arbitrary extraterrestrial transmitter is unknown, and thus Doppler drift represents a search parameter for narrow-band SETI. In a common data dumping time, say 1s, the implied Doppler drift will be much smaller than the broadening of any known interstellar spectral lines, thus necessitating the fine spectral resolution of SETI.

FAST is collaborating with the \BL\ group, who has developed an efficient narrow-band search software package which includes a search for such drifting signals, named \tseti\footnote{\url{turboSETI: https://github.com/UCBerkeleySETI/turbo_seti}}.
For the proposed SETI campaign with FAST, we will use \tseti\ to conduct a similar search and candidate selection procedure as described in \cite{2017ApJ...849..104E}.

\subsection{Broad-band signals and modulation classification using machine learning}
Traditionally, radio SETI has focused on searches for narrow-band signals. In this section,
we highlight some of the newly-developed tools that have as yet not been fully explored for SETI.
Along with narrow-band signals, we plan to conduct searches on two different types of broad-band
signals; wide-band artificially dispersed pulses and signals exhibiting artificial modulation.

\subsubsection{Artificially-dispersed pulses}

\begin{figure}[h]
    \centering
   \includegraphics[scale=0.15, angle=-90]{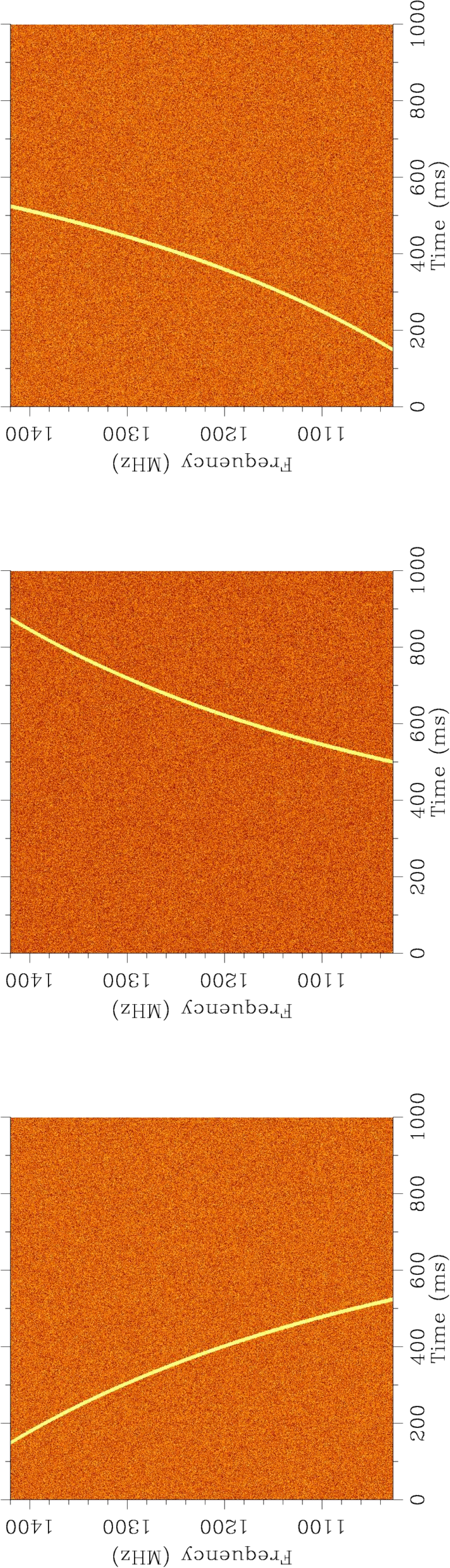}
    \caption{Three negatively-dispersed simulated pulses. The newly developed pipeline which utilizes machine learning (ML) techniques will be capable of searching these type of dispersed pulses which the standard transient technique is unlikely to investigate.}
    \label{fig:negativeDM}
\end{figure}

Astrophysical sources such as pulsars \citep{Hewish:1968}, rotating radio transients \citep{mclaughlin:2006}
and fast radio bursts \citep{Lorimer:2007p5652} exhibit broad-band pulses that are dispersed due to their propagation through the intervening ionized medium. This dispersion causes the higher-frequency component of the pulse to arrive earlier than the lower-frequency component. FAST has already demonstrated its ability to find such signals by discovering around 70 new pulsars in less than a year, \citep{Qian2019}\footnote{\url{http://crafts.bao.ac.cn/pulsar/}} including finding a pulsar with interesting emission properties \citep{2019arXiv190405482Z}. \cite{Siemion:2010p6845} have speculated an interesting hypothesis that an advanced civilization might intentionally create a beacon of ``pulses" with artificial (nonphysical) dispersion. In addition, they also suggested that the energy requirement for such a signal is relatively similar to the energy required for a persistant narrow-band signal. There have been a few attempts to search for such signals \citep{vonKorff:2010p3275,Harp:2018apj}; however, no detailed investigations have been carried out.
Figure \ref{fig:negativeDM} shows examples of three different types of artificially-dispersed engineered signals which are clearly distinguishable from naturally-occurring dispersed pulses. The \BL\ team have developed tools to search for these classes of signals (\citealt{zhang2018FRB}). With the excellent sensitivity of FAST, the aforementioned targeted searches would be ideal to investigate such signals.

\subsubsection{Modulating signals of extra-terrestrial original}
\begin{figure}[h]
    \centering
    \includegraphics[scale=0.3,angle=-90]{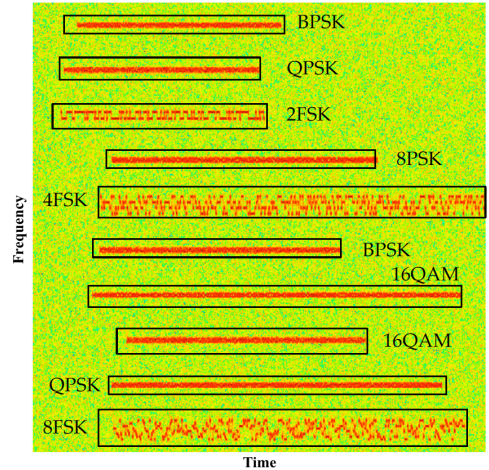}
    \caption{Example of seven artificially generated modulated signal embedded in wide radio band. The plot shows observed radio frequency (RF) as a function of time in arbitrary units (credit: \citealt{zpq+19})}
    \label{fig:example_waterfall_10modulation}
\end{figure}{}

Modulation schemes are methods of encoding information onto high-frequency carrier waves, making the transmission of that information more efficient. Most of these methods modulate the amplitude, frequency, and/or phase of the carrier wave. Broad-band radio emissions exhibiting such underlying modulation represent a third important class of radio emission indicative of an artificial origin, as we would expect any transmission containing meaningful information to exhibit some form of modulation. \cite{2015arXiv150600055H} carried out a simple modulation signal search using correlation statistics towards known astrophysical sources such as pulsars, quasars, supernova remnants, and masers. In the last few years, great progress in the field of machine learning (ML) has opened up myriad new opportunities in this area. Moreover, Convolutional Neural Network (CNN) classifiers, heavily used in computer vision applications, provide advanced capabilities with the aid of high-performance computing. In CNNs, classification of an input signal is carried out through alternating convolution and pooling layers along with a final fully--connected layer providing the desired output classes (see Figure 1 in \cite{CNN_arch} and references therein). Convolution layers are trained using labeled data of desired classes to extract desired features from a given input signal. After training, the network learns local features to map a given input signal to its closest output class by minimizing a loss function \citep{CNN_arch}.

\begin{figure}[h]
    \centering
    \includegraphics[scale=0.3]{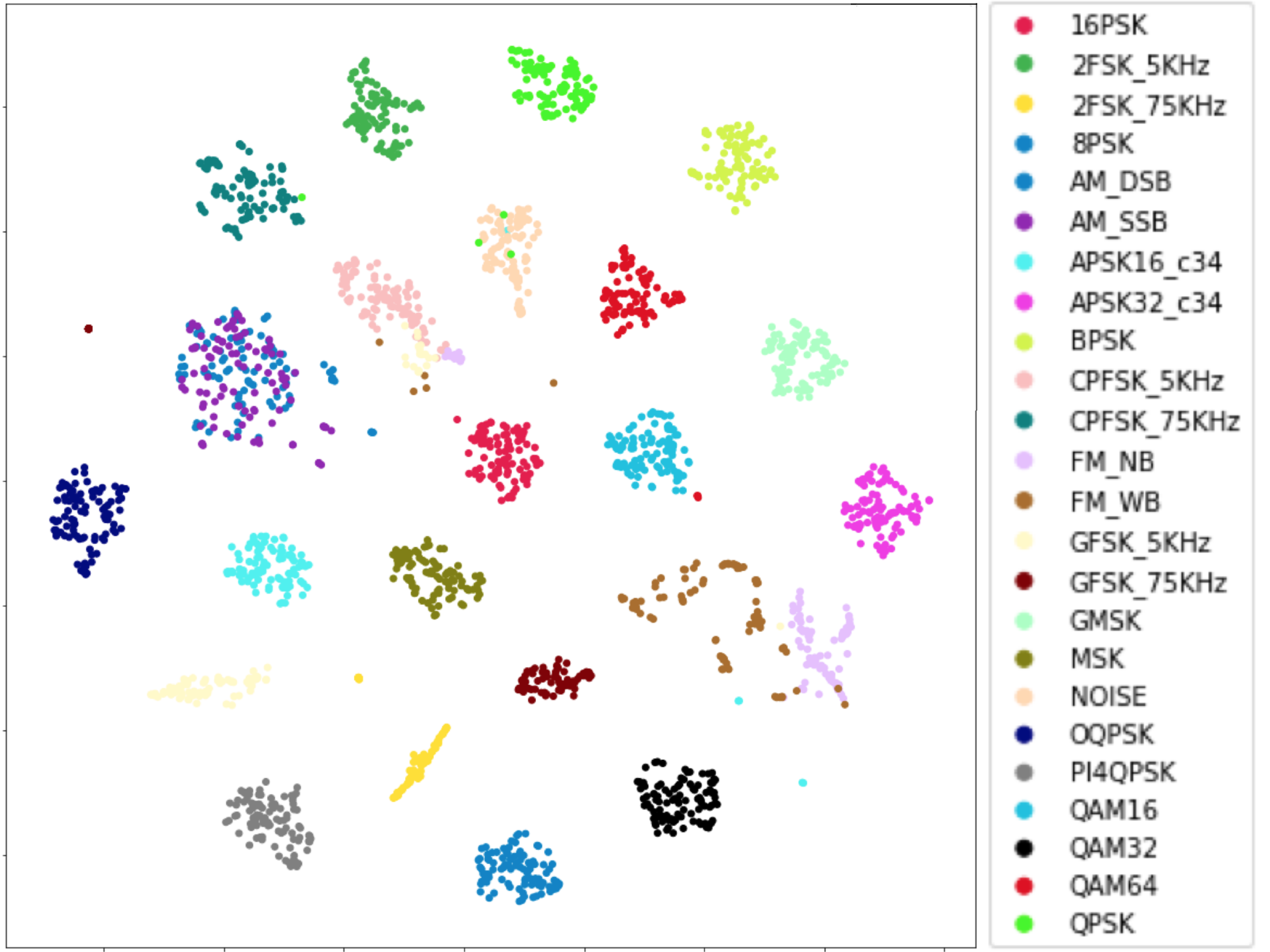}
    \caption{Application of a modulation classifier developed by the Breakthrough Listen team on the simulated radio-frequency data with embedded modulations. The figure shows the T-distributed Stochastic Neighbor Embedding (t-SNE) plot embedding 24 modulations using one of our models. The axes here are arbitrary as they merely represent the space found by t-SNE in which close points in high dimension stay close in the lower dimension. As it is shown, different modulations map to different clusters even in 2-dimensional space indicating that the model does well in extracting features that are specific to the different modulation schemes (Zhang et al.\ 2020 in prep)}
    \label{fig:ml_class}
\end{figure}

Recently, there have been ongoing efforts to classify various modulation types using CNNs and Deep Neural Networks (DNNs) for real-world applications (see \cite{DeepSig_application} and \cite{zpq+19} for detailed discussions). Figure \ref{fig:example_waterfall_10modulation} shows one such example of ten artificially generated modulation signals across a wide range of frequencies from \cite{zpq+19}. Moreover, a US-based startup, {\itshape DeepSig}\footnote{\url{www.deepsig.io} have provided labeled real-world RF spectrum datasets\footnote{\url{www.deepsig.io/datasets/}} of around 24 modulation types for ML algorithm development. 
Such complex signals, although anthropogenic in nature, constitute a completely new class of signal which has never been comprehensively searched for in SETI applications.} Members of the \BL\ team have utilized datasets provided by the US Army Rapid Capabilities office’s Artificial Intelligence Signal Classification challenge. In the challenge, labeled datasets containing 24 modulation classes were provided with six degrees of signal-to-noise ratio (SNR). The \BL\ team was successful in developing a CNN based classifier to achieve 95\% prediction accuracy for the high SNR signals\footnote{\url{https://github.com/moradshefa/ml_signal_modulation_classification}}. A t-SNE plot embedding these 24 modulation signals with well-separated clusters in 2-dimensional space is shown in Figure \ref{fig:ml_class}.
%A confusion matrix for this similar classification for the high SNR signals is shown in Figure \ref{fig:confusion_matrix}. 

%\begin{figure}[h]
%    \centering
%    \includegraphics[scale=0.3]{Confusion_matrix_CNN_modulation.pdf}
%    \caption{Confusion matrix between the predicted and true modulation classes %for the high SNR signals with the color-bar showing prediction accuracy. 
%    As it is apparent, our classification was able to predict correct % %modulation types with high accuracy for all 24 modulation signal types.}
%    \label{fig:confusion_matrix}
%\end{figure}

%
\begin{figure}[h]
    \centering
    \includegraphics[scale=0.2]{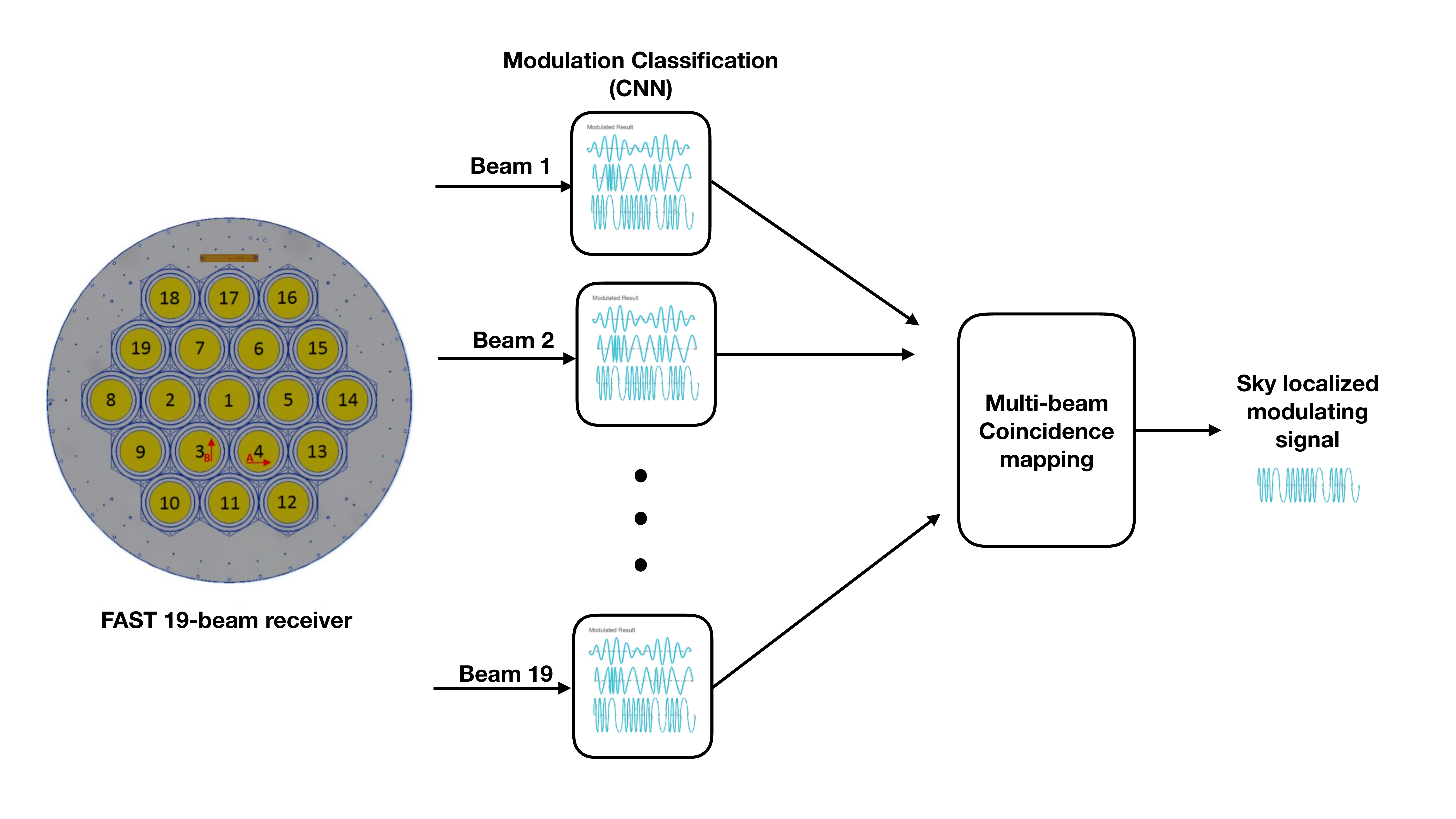}
    \caption{A schematic of a multi-beam modulation classification and the localizing scheme proposed with the FLAN receiver. The first block represents the standard CNN based modulation pattern identification possible to carry out in the individual dedicate compute node. The second block -- running on the headnode or one of the extra compute node -- receives time-stamp, modulation type, range of signal frequencies, and SNR for every candidate from each beam for a multi-beam coincidence mapping. Any signal with sufficient SNR and only found in 4 nearby beams with similar characteristics can be considered a sky localized candidate.}
    \label{fig:modulation_multibeam_coinc}
\end{figure}

%For example, classification of the modulation format of received signals (in realtime for some cases) is a necessary task for the modern intelligence community. These algorithms are required to work regardless of a prior knowledge of the received signals, such as power, frequency, and phase, and must be coupled with signal degradation and interference.

We plan to deploy a similar signal modulation analysis pipeline using energy detection and modulation classification via a CNN classifier in collaboration with the \BL\ team. Such signal searches require recording of baseband raw voltages which would be possible to carry out with the proposed computing cluster in collaboration with the \BL\ team (see Figure \ref{fig:BL_backend_nodes}). One of the challenges of these searches is that a large fraction of the RFI tends to utilize one of a limited number of modulation schemes. However, sky localized modulation patterns of known and unknown types are of great interest to SETI. The FLAN provides a unique opportunity to scrutinize such signals using its coverage from 19 independent sky pointings with a significantly small overlap. Anthropogenic signals are likely to appear in multiple beams while a real sky localized signal -- originating from a point source -- can cover three to four nearby beams. A schematic of such a pipeline is shown in Figure \ref{fig:modulation_multibeam_coinc}. Such multi-beam coincidence mappings have been successfully used in searches for the FRBs at other telescopes with multi-beam receivers (see for example \citealt{pfg+19}). For FRBs, a comparison of detected time-stamp, dispersion measure, and SNR can be carried out across all beams for every candidate. Similarly, for every detected modulation signal, corresponding time-stamp, modulation type, range of signal frequencies, and SNR can be compared across all beams. We plan to use such a technique to significantly reduce the number of false positives for such modulation signal searches with FAST. 

Thus such techniques could allow a sensitive search with FAST for modulated signals from ETIs towards the targets mentioned in Section \ref{sect:setiWfast}. Furthermore, the \BL\ team has also developed CNN based ML classifiers to detect narrow-band \citep{Zhang:2018ue} and dispersed pulses \citep{zhang2018FRB} which are also possible to deploy.

%As shown, the classifier was able to successfully differentiate between all 24 different modulation types (Zhang et al. 2019 in prep).
%On Earth, the bandwidth of modulated signals varies from a few kHz\footnote{\url{https://www.mwrf.com/markets/understanding-ultra-narrowband-modulation}} to many MHz\footnote{Some of the LTE transmissions do have such a large bandwidths}, based on the type of modulation.

\section{Sensitivity and Rarity of ETI transmitter}
\label{sect:limits}
The required power for a certain ETI transmission to be detected depends on its directionality and other characteristics of the signals.  We thus introduce the effective isotropic radiated power (EIRP;  \citealt{2017ApJ...849..104E})  as
\begin{equation}
    EIRP = \sigma \frac{4 \pi d_\star^2} {A_{eff}^{R}} \frac{2 k T_{sys}} {\sqrt{ n_{p}  t_{obs}  \delta{\nu}} }  \>\>,
    \label{Eq:EIRP}
\end{equation}
where  $\sigma$ is the required S/N, $\delta{\nu}$ is the bandwidth of the transmitted signal, $t_{obs}$ is the observing integration time, $A_{eff}^{R}$ is the effective aperture of the receiver on Earth, $n_p$ is the number of polarization, and $d_\star$ is the distance between the transmitter and the receiver, i.e., distance to the star.

It is straightforward to estimate the required transmitting power to be detected for any source at a certain distance.
\cite{barclay:2018} estimated that the median distance of stars with potential exoplanets that TESS will find to be around 200 light-years. To a FAST-equivalent system of 300 meter aperture with a 70\% antenna efficiency, the required EIRP 
%\del{to detect signals above 0.6 Jy} 
is of the order of 1.9$\times$10$^{11}\ $W with a 1 Hz channel with 5-minutes integration. Similarly, we also estimated EIRP limits at other wavebands in Table \ref{tab:all_receivers}. It should be noted that humans routinely produce planetary radar signals with EIRP of the order of $\sim$10$^{13}\ $W, 
%\del{However, One could easily envision a situation in which a marginally more advanced civilization than our own might possess a significantly more powerful system used for a similar space surveillance application. Moreover, using a higher gain antenna with a similar transmitter power, it would be possible to achieve higher EIRP of the order of $10^{12}\ $ to $10^{13}\ $W} 
which is higher than EIRP limits at all wavebands in Table \ref{tab:all_receivers}. Thus, FAST will be able to put tighter constrains on any putative narrow-band signals towards stars with newly discovered exoplanets in the solar neighbourhood.

For the survey of the Andromeda galaxy, as shown in Figure \ref{fig:FAST19beam}, we can easily cover the entire galaxy using just 21 hexagon patterns, which corresponds to 84 pointings. With 10 minutes of 
%\del{one hour} 
integration time per pointing, FAST will be able to detect an EIRP of 2.4
%\del{1.3}
$\times$10$^{19}$ W at the 0.77 Mpc distance of Andromeda. 

%\del{we can estimate that the minimum detectable flux will be of the order of 0.37 Jy. This will allow us to calculate EIRP of a possible narrow-band transmitter in the Andromeda galaxy as}
%\begin{equation}
 %   EIRP_{Andromeda}~=~0.37\times{10^{-26}}\times4\pi{d^2} ~ Watts.
  %  \label{eq:andromeda}
%\end{equation}
%\del{Here, d is the distance of the Andromeda galaxy which is roughly around 2.4$\times$10$^{22}$ meters. Putting this in Equation \ref{eq:andromeda}\, we can estimate that the EIRP limit is around 2.4$\times$10$^{19}$ W. } 

This is three orders of magnitude higher than the total energy budget of a Kardashev Type I civilization ($\sim$10$^{17}$ W) but significantly lower than the energy budget of Kardashev Type II ($\sim$10$^{26}$ W) and Type III ($\sim$10$^{36}$ W) civilizations. Thus, such a dedicated survey will be able to put tight constraints on the presence of a transmitting Type II and Type III civilization among the 1 trillion stars of the Andromeda galaxy 
%\del{in just 6 hours of observing with FAST}.

\subsection{Rarity of ETI Transmission limit comparison with other SETI surveys}
\begin{figure}[h]
    \centering
    \includegraphics[scale=0.3]{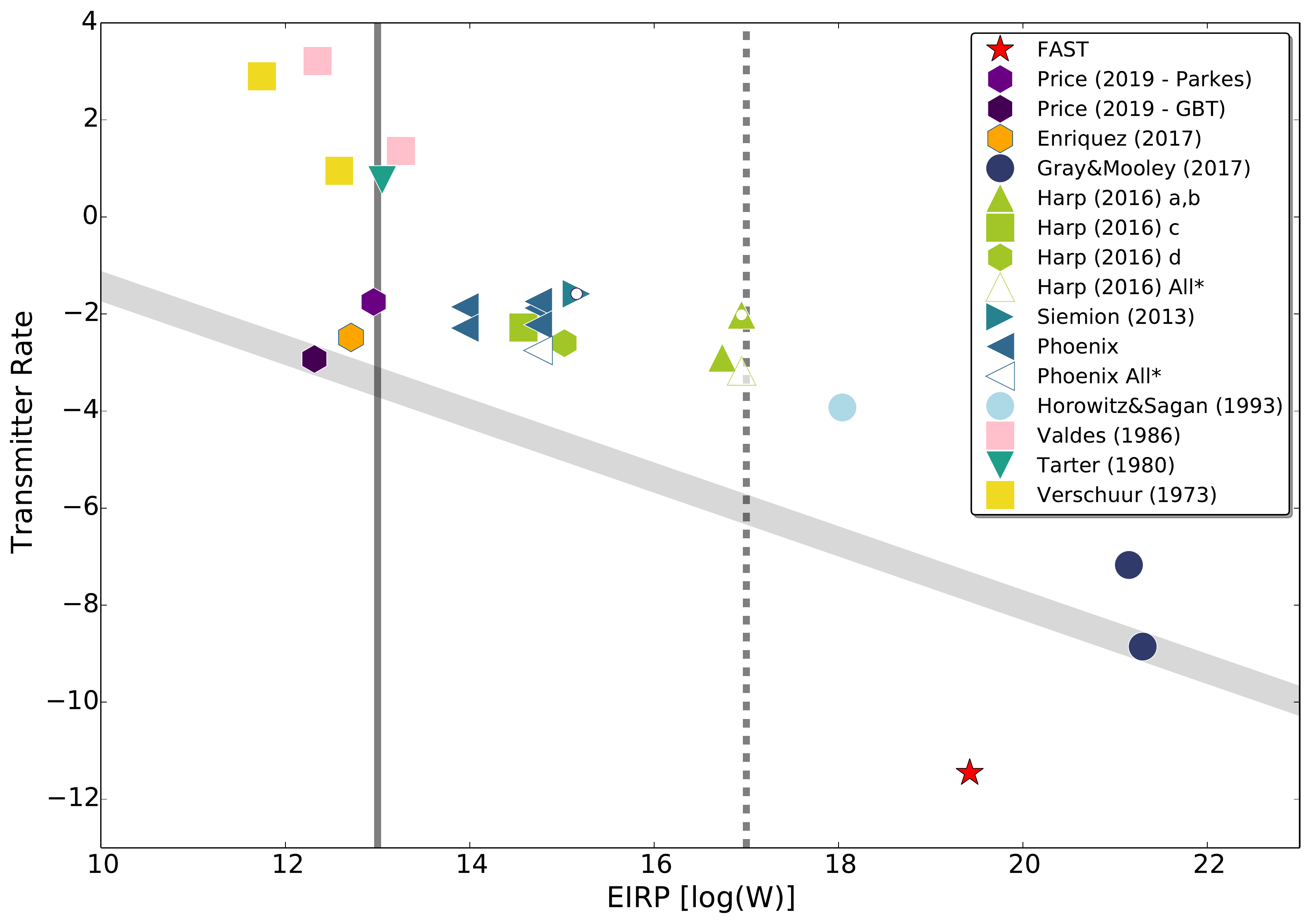}
    \caption{Transmitter rate (or rarity of ETI transmitter) vs EIRP limit for a survey of the Andromeda galaxy with FAST compared to all the previous significant SETI surveys. The solid line presents the EIRP of modern day narrow-band planetary radars, the dotted line is the speculative EIRP of a Kardashev Type I civilization. The filled region shows current limits placed by some of the sensitive modern-day surveys. The red-star shows the expected transmitter rate limit with a 14-hour observing campaign of the Andromeda galaxy using FAST's 19-beam receiver, which achieves much better sensitivity than the current limits (see  \cite{2017ApJ...849..104E} and  \cite{2020AJ} and references therein for a more detailed discussion).}
    \label{fig:EIRPlimits}
\end{figure}

As mentioned in Section \ref{sect:intro}, over the past 60-years several SETI surveys have been carried out using a number of radio telescopes. These surveys include targeted searches towards nearby stars and nearby galaxies, and blind surveys of the sky. \cite{2017ApJ...849..104E} have suggested a quantitative comparison parameter to compare these different SETI surveys. The rarity of ETI transmitters or transmitter rate is
one of a possible way to compare these surveys. This parameter can be given as,
\begin{equation}
{\textit{Transmitter~rate}}~=~log \Big(\frac{1}{N_{stars} \nu_{relative}} \Big).
\end{equation}
Here, N$_{stars}$ is the number of stars surveyed and $\nu_{relative}$ is the relative bandwidth of radio spectrum covered. Figure \ref{fig:EIRPlimits} shows the transmitter rate as a function of sensitivity for these surveys. It can be seen that 14-hour (84 pointing $\times$ 10-minutes) survey of the Andromeda galaxy, using the FAST's 19-beam receiver, will be well below the most sensitive limits placed by all the earlier SETI experiments.

{}

%The multiyear CRAFT survey will search for pulsars, fast radio bursts, radio signals from extraterrestrial civilizations, as well as map galactic and extragalactic hydrogen. $(Di Li et al,  https://arxiv.org/abs/1802.03709)

%The FAST SetiBurst instrument is designed to perform two of these five searches - Fast Radio Bursts and SETI signals.

%1. Commensal SETI backend and Berkeley collaboration.
%2. BL-SETI MOU

\section{Summary}
\label{sect:summary}
FAST is the largest single-aperture radio telescope in the world and provides unprecedented sensitivity.  In a collaboration with \BL, we have equipped the FLAN multi-beam receiver of FAST with a SETI pipeline, namely,  {\itshape FAST-SetiBurst}, which has been tested targeting GJ273b, one of the closest known exoplanets and placed an EIRP detection limit of 7$\times$10$^{10}$ W. Based on these characterization of the FAST SETI systems, we outline two unique FAST SETI experiments, both of which will push the current limits placed by earlier studies. First, a survey of the Andromeda galaxy (M31) will detect an ETI with EIRP $\textgreater$ 2.4$\times$10$^{19}$  W, corresponding to a comprehensive coverage of Kardashev type II and III civilizations in an external galaxy.
Second, a survey of TESS exoplanets will detect an ETI with EIRP $\textgreater$ 1.9 $\times$10$^{11}$  W, which is well within the reach of current human technology.  Along with narrow-band signal searches, we will also deploy comprehensive searches for artificially-dispersed pulses and modulated signals towards all targets. FAST will place meaningful limits on possible ETI transmitters coincident with these targets and help answer humanity's oldest question: are we alone?

\bibliographystyle{raa}
\bibliography{references}

\end{document}